# Novel Silicon-Carbon Fullerene-Like Cages: A Class of $sp^3 – sp^2$ Covalent-Ionic Hybridized Nanosystems

M. N. Huda and A. K. Ray*
Department of Physics
The University of Texas at Arlington
Arlington, Texas 76019

## Abstract

A class of highly symmetric silicon carbide fullerene-like cage nanoclusters with carbon atoms inside the $Si_{20}$ cage and with high stability are presented. The Generalized Gradient Approximation of Density Functional Theory (GGA-DFT) is used to study the electronic and geometric structure properties of these structures and full geometry optimizations are performed with an all electron 6-311G** basis set. The stability of the clusters is found to depend on the geometrical arrangements of the carbon atoms inside the clusters and the partly ionic nature of the bonding. Possibilities of extending these structures into a larger class of nanostructures are discussed.

Clusters are distinctly different from their bulk state and exhibit many specific properties, which distinguishes their studies as a completely different branch of science named "Cluster Science". Large surface to volume ratio and quantum effects resulting from small dimensions are usually prominent in clusters and ideas like 'super-atoms', 'magic numbers' or 'fission' in clusters [1-4] have prompted a wide class of scientists to study this 'relatively' new area of the physical sciences. Cage-like compact clusters are particularly important for two reasons: they can be used as building blocks of more stable materials and the hollow space inside the cage can be used to dope different atoms yielding a wide variety of atomically engineered materials. For example, well-controlled nanostructures with varying HOMO-LUMO gaps and desired conduction properties can be achieved by controlled doping of atoms in $C_{60}$.[5] The spin property of the doped atom inside the cage can be used as the smallest memory devices for quantum computers; for instance, tungsten in $Si_{12}$ clusters is quantum mechanically isolated from outside so that it can preserve its spin state.[6]

Silicon is one of the more important semiconductors with widespread applications and silicon clusters, preferring $sp^3$ hybridization, have been studied extensively. Hartree-Fock (HF) and density functional theories (DFT) [7-9] have been used to determine the ground state structures, though still controversial, of Si clusters. There are not, however, enough experimental studies to confirm or predict the energetically favorable structures.[10] Discovery of the magically stable $C_{60}$ fullerene cage have prompted scientists to study fullerene-like silicon structures and $Si_{60}$ was found to have a distorted fullerene-cage-like structure.[11] Attempt also has been made to replace carbon atoms by Si atoms in $C_{60}$, also resulting in a distorted structure.[12] In carbon clusters, preferring $sp^2$ hybridization, fullerene like structures are found in structures as small as $C_{20}$.[13-14] In $Si_n$ clusters such structures are unstable for small n values.

---

*email: akr@exchange.uta.edu

It has been pointed out recently, primarily based on DFT studies, that highly stable small silicon cage clusters are possible if transition *metal* atoms are encapsulated in the cage.[15-18] The combinations of silicon and carbon atoms in a cluster have also generated a number of studies on structures *rich in carbon atoms*, in areas from cluster science[19] to astrophysics[20]. Density functional calculations with simulated annealing have been performed by Hunsicker and Jones[21] for neutral and singly charged silicon-carbon cluster anions with up to eight atoms. The calculations identified two classes of anion structures: carbon-rich (chainlike) and silicon- rich (three-dimensional), with pronounced differences in the vertical detachment energies. The largest silicon-rich cluster studied was $Si_7C^-$. Photolysis experiments on SiC mixed clusters have been reported by Pellarin *et al.*[22] The experiments indicate surprising capability of substituting a large number of silicon atoms, up to 12, into fullerenes without destabilizing their cage structure significantly.

However, to the best of our knowledge, *Si rich cage type* silicon carbide clusters have not been studied in detail so far. We have recently shown that *carbon dimers* trapped into medium size silicon clusters produces structures *comparable in stability* to metal encapsulated silicon cage clusters. For this purpose, we carried out *ab initio* Hartree-Fock based second order Møller Plesset perturbation theory calculations to study the electronic and geometric structures of $Si_nC_2$ (n = 8-14) clusters and predicted $Si_{14}C_2$, with a close fullerene like structure, to be a magic cluster.[23] This theory was recently applied by us to predict that $Ag_8$ is a magic cluster.[24] Given the similarities between carbon and silicon and the fact that $C_{20}$ can make the smallest fullerene system, we report here a novel class of $Si_{20}C_n$ systems. It is well known that bare silicon clusters do not form closed structures, because of their $sp^3$ bonding nature. $Si_{20}$, for example, is a prolate type structure with two $Si_{10}$ units joined by intermediate bonds[25]. We demonstrate here, with gradient corrected density functional theory (DFT) [26] and an all electron 6-311G** set[8], that *multiple carbon atoms inside the $Si_{20}$ do in fact produce highly stable fullerene like structures.* The GAUSSIAN 98 suite of programs[27] has been used.

Figure 1 shows the $Si_{20}C_n$ ($3 \leq n \leq 6$) optimized structures and table 1 lists the binding energies per atom (BE), highest occupied molecular orbital – lowest unoccupied molecular orbital (HOMO-LUMO) gaps, vertical ionization potentials (VIP) and Vertical electron affinities (VEA). The binding energies per atom of the clusters are computed as the relative energies of the clusters in the separated atom limit, with the atoms in their respective ground states. The VIP and the VEA are calculated as the difference in total energies between the neutral clusters and the corresponding positively and negatively charged clusters, respectively, at the neutral optimized geometry. We have considered different numbers of carbon atoms with various possible orientations in the Si cages and the structures are divided into four different categories. We wish to emphasize here that given the large number of atoms in the cluster, the number of possible structures are quite large. Only the converged, most stable structures are reported here. For the first set, the input geometry was a silicon cage with pentagons (Penta-*n*) and in table 1 or figure 1, 'Penta-4', for example, means that the input geometry was a pentagonal cage with four carbon atoms inside it. The second category is a cubic (Cube-*n*) cage, where the cubes are placed on top of another cube to form a chain-like cage with the carbon atoms inside the cage. This is followed by a hexagonal cage (Hex-*n*), where the cages are made up of Si hexagons, with carbon atoms inside. The last one is the octagonal cage (Octa-*n*) where Si octagons are used to form the cages. The minimum number of carbon atoms inside the cage reported here is three and the maximum number is six. In our previous study, only carbon dimers in $Si_8$ to $Si_{14}$ clusters were considered.[23] However, as the cage size increased to twenty silicon atoms, we found that at least three carbon atoms were necessary to stabilize the cage.

We have four structures in the Penta-series as shown in figure 1. Penta-6 is more like a prolate structure with a BE per atom of 4.16eV. Out of six carbon atoms, four formed a rhombus-shaped structure at the middle, while the other two are near the two rear ends, below the two Si capping atoms. Penta-5 has the same BE per atom as Penta-6, but is more compact. Here three carbon atoms formed a chain in the middle, and the other two carbon atoms are at the boundary of the cage. This cluster is a unique cluster in that the carbon atoms are on the surface of the cage. However, both the VIP and VEA are higher than the Penta-6 cluster, indicating increased stability. The HOMO-LUMO gaps are comparable. In Penta-4, the four carbon atoms again form a rhombus in the middle. It has the lowest binding energy in the Penta-series, 3.92eV per atom. Penta-3, where all three carbon atoms formed a linear structure in the middle, has a high binding energy of 4.01eV per atom and the highest VIP and VEA in this series.

In the cubic series, the structures are cubic chain like structures. For Cubic-5, each of the rectangular layers has one carbon atom each; with each rear end carbon atoms being clearly inside the cage as shown. Binding energy is the highest in the series, 4.05eV per atom. We believe that longer cubic chains, *similar to carbon nanotubes,* can easily be constructed by extending this structure on both sides. In Cubic-4, carbon atoms are in the two rear end cube, leaving the middle part empty. Among the four-carbon atom clusters, Cubic-4 is the most stable. In Cubic-3, the three carbon atoms formed a linear chain in the middle part of the cage, leaving the two outer cubes empty. Cubic-3 has the lowest binding energy of 3.20eV per atom among the class of clusters considered in this study, possibly due to the linear arrangement of the carbons. It also has the lowest HOMO-LUMO gap of 0.16eV. For this series, HOMO and HOMO-1 are found to be degenerate.

The optimized structures of Hexa-5, is distorted from the straight hexagonal cage, converging to the capping atoms at both ends. Out of five carbon atoms, one is in the center and the other four are like the inside capping of the four walls of the cage. The binding energy of 3.94eV per atom and a HOMO-LUMO gap of 0.32eV are lower than the previous two five-carbon atom clusters. Hexa-3 is like a straight hexagonal cage with the three carbon atoms along the axis of the hexagons. The binding energy of 3.72eV per atom is lower, again possibly due to the linear arrangement of the carbons and the gap is the same as the gap of Hexa-5.

The last set is the octa-series consisting of two octagons capped with two Si atoms on the top and two at the bottom. For this series we have three optimized structures as shown in Figure 1. Octa-6 and Octa-4 has almost the same structures with Octa-6 being slightly more circular. Octa-6 has a binding energy of 4.12eV per atom and a HOMO-LUMO gap of 0.65eV, comparable to the Penta-6, though its VIP and VEA are slightly higher. Octa-4 has a binding energy of 3.90eV per atom, and a lower HOMO-LUMO gap of 0.34eV. Its VEA is one of the highest. Octa-3 has a different structure than the other two in this series. The structure does not appear to be circular and the three carbon atoms sit in the cage as a triangle. The binding energy is 3.83eV per atom; and the gap is 0.18eV, comparable to the Cubic-3 gap.

From Table 1 it is evident that binding energy per atom increases with the number of carbon atoms in each group. Only one exception was found for the pentagonal group where Penta-3 is more stable than the Penta-4. The compactness of Penta-3 might contribute to its higher binding energy. Also the two six-carbon atoms structures Penta-6 and Octa-6 have almost similar binding energies. We believe that the coordination of carbon atoms to silicon atoms contributes to the binding energy. Closer the number of silicon atoms to the number of carbon atoms, more bound is the system. Mulliken charge distribution analysis indicates that in most cases, carbon atoms acquired negative charges and the silicon atoms acquired positive charges, as is also expected from electronegativity considerations. Two exceptions are noted: one is in Penta-3 where the middle carbon atom gets positive charge (+0.99$e$), but the other two carbon atoms got −1.25$e$ charge each. Another is in Penta-5 where in the middle three carbon atoms, the center one acquired +1.27$e$ and the two end atoms got −1.37$e$ each. This means that there is a strong Columbic interaction between the carbon atoms contributing to their high binding energies. Since normally Si atoms participate in $sp^3$ bonding and carbon atoms prefer $sp^2$ bonding, we have here a hybrid $sp^3$ and $sp^2$ bonding, with more contribution from $sp^3$ as the number of Si atoms is increased. Distortions in pure silicon clusters are also believed to be due to $sp^3$ bonding. We believe that the admixture of the two types of bonding, *covalent and ionic,* helps minimize the distortions on the cluster surfaces and contributes to fullerene like structures.

Vertical ionization potentials are considerably higher for all the clusters discussed above. The highest one of 6.96eV is for Cubic-4, while the lowest one is for the Hexa-3 (5.87eV). No general pattern is however observed in the ionization potentials. They do not depend on the number of carbon atoms in the clusters; the particular structures seem to influence these potentials. Similar comments are found to be true for the vertical electron affinities. These EAs are similar to the DFT study of chromium encapsulated $Si_{11}$ to $Si_{14}$ clusters.[16] Also IPs are of the same order. Form Table 1 we can also conclude that, in general, the HOMO-LUMO gaps increase with the carbon atoms. Also, except for Octa-3 (dipole moment of 0.176D), all the other structures have zero dipole and quadrupole moments. This implies that the overall charge distribution is symmetric contributing to their stability.

As indicated before, we find that putting at least three carbon atoms inside the silicon cage helps symmetrize and stabilize the Si cage. For example, placing a carbon dimer along the axis in the middle of a cubic chain (three cubes put one on another) with 20 silicon atoms did not stabilize it. However putting three atoms at the center along the axis not only made it a perfect cubic chain but the binding energy per atom also increased. Similar situations were observed for the other groups. In a recent study of $Si_{60}$ with $C_{60}$ fullerene inside the silicon cage, it was observed that the overall structures were highly distorted.[28] Our study along with this fact suggests that the number of carbon atoms in the cage is a variable, which has to be optimized with respect to the number of silicon atoms on the cage surface to yield highly symmetric and stable cages. If we increase the silicon cage size, the number of carbon atoms inside the cage, their orientations and geometries haves to be determined carefully.

In conclusion, we have studied a class of highly symmetric and highly stable $Si_{20}C_n$ clusters. The stability is found to depend on the number of carbon atoms in the $Si_{20}$ cage as also their orientations. The ionic part of the bonding plays a major role in the electronic and geometric properties of these clusters. Further theoretical and experimental study are urgently needed to give precise information about hybrid bonding, symmetry and stability of larger cages than considered here.

Finally, the authors gratefully acknowledge partial support from the Welch Foundation, Houston, Texas (Grant No. Y-1525).

Table 1: Binding energy (B.E.) per atom, HOMO-LUMO gap, VIP and VEA (all in eV) for optimized $Si_{20}C_n$ clusters.

| Structure | State | B.E./atom | Gap | VIP | VEA |
|---|---|---|---|---|---|
| Penta-3 | $^1A_g$ | 4.01 | 0.56 | 6.60 | 3.09 |
| Penta-4 | $^1A_g$ | 3.92 | 0.59 | 6.42 | 2.77 |
| Penta-5 | $^1A_g$ | 4.16 | 0.70 | 6.52 | 2.89 |
| Penta-6 | $^1A_g$ | 4.16 | 0.66 | 6.10 | 2.47 |
| Cubic-3 | $^1A_{1g}$ | 3.20 | 0.16 | 6.15 | 3.15 |
| Cubic-4 | $^1A_{1g}$ | 3.98 | 0.32 | 6.96 | 3.13 |
| Cubic-5 | $^1A_{1g}$ | 4.05 | 0.57 | 6.27 | 2.84 |
| Hexa-3 | $^1A_{1g}$ | 3.72 | 0.32 | 5.87 | 2.59 |
| Hexa-5 | $^1A_g$ | 3.94 | 0.32 | 6.50 | 3.18 |
| Octa-3 | $^1A_1$ | 3.83 | 0.18 | 6.27 | 3.13 |
| Octa-4 | $^1A_g$ | 3.90 | 0.34 | 6.53 | 3.18 |
| Octa-6 | $^1A_g$ | 4.12 | 0.65 | 6.56 | 2.98 |

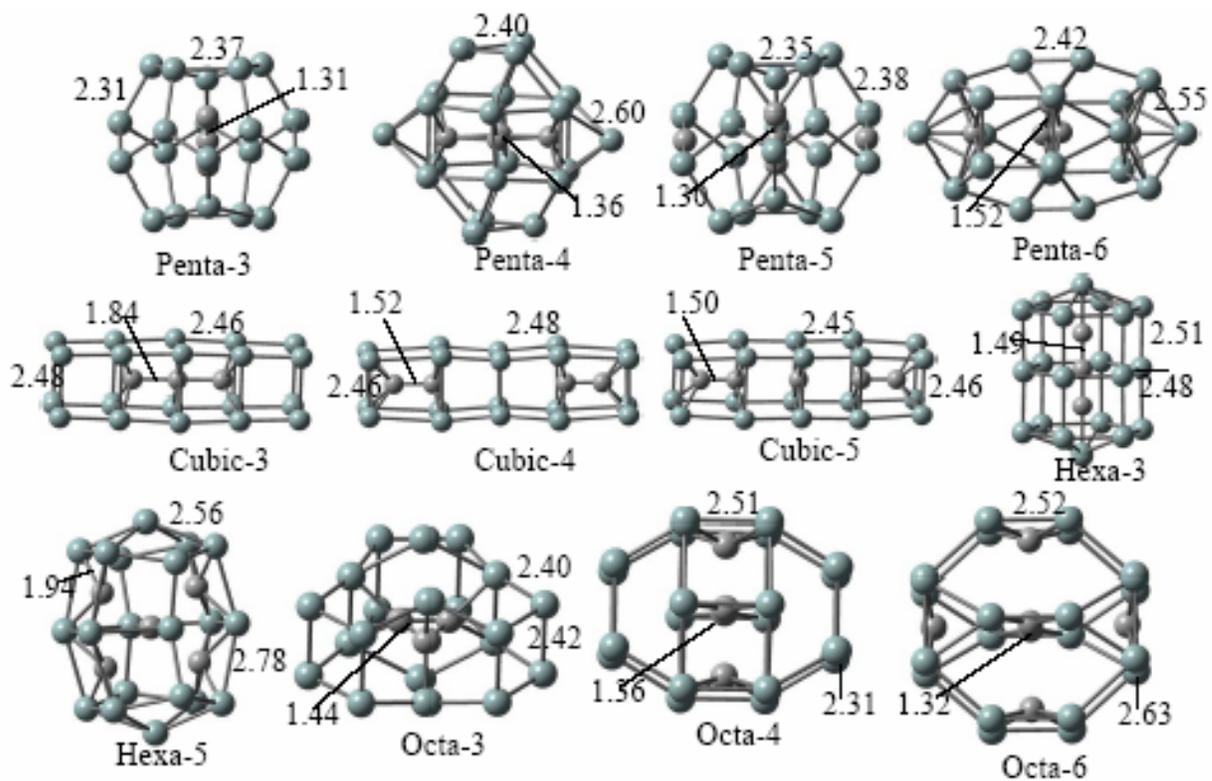

Figure 1: Optimized structures of Si20Cn clusters.

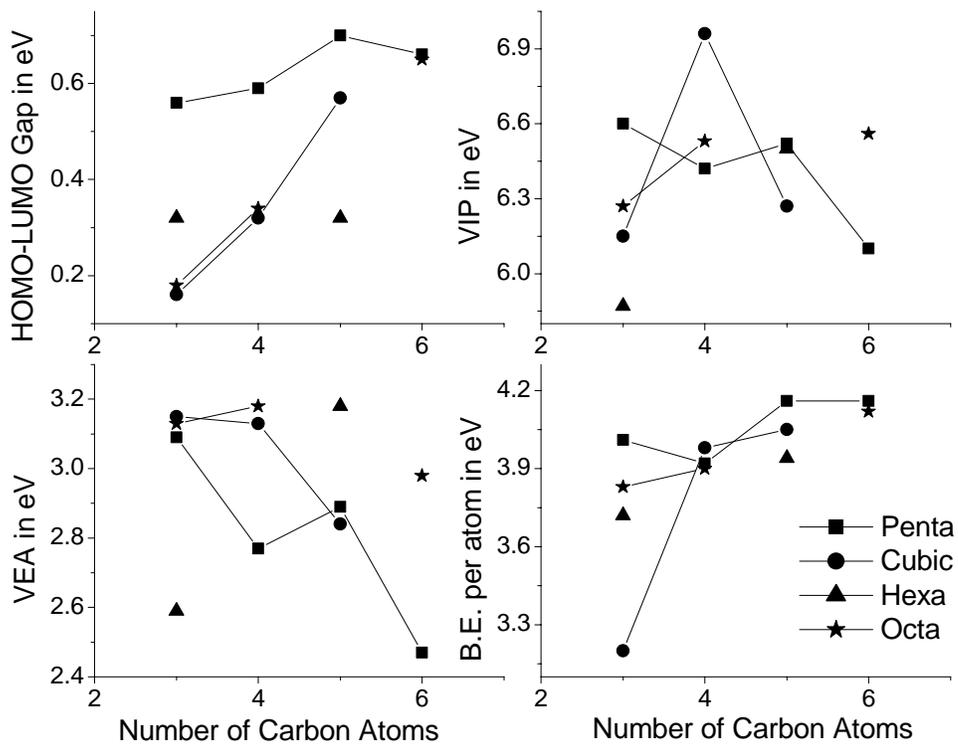

Figure 2: HOMO-LUMO gap, VIP, VEA and B.E. per atom in eV versus the number of carbon atoms in the $Si_{20}$ nano-structures.